\documentclass[10pt]{iopart}
\begin{document}

\title[]{ 
Highlights of Cern Workshop on \\
~Charm Production in A+A Collisions }

\author{Marek Ga\'zdzicki \dag  
\footnote[3]{
Marek.Gazdzicki@cern.ch}
}

\address{\dag\ CERN, Geneva, Switzerland, \\
Institut f\"ur Kernphysik, Universit\"at Frankfurt, Frankfurt, Germany}

\vspace{1cm}

\begin{abstract}
Models and experimental effort concerning open and
hidden charm production in nuclear collisions
discussed at Cern Workshop in December 99 are reviewed.
The most recent development is also mentioned.
\end{abstract}


\vspace{4cm}
\begin{center}
{\it Invited presentation at 5th International Conference on \\
Strangeness in Quark Matter, \\
Berkeley, California, July 20--25, 2000}
\end{center}

\maketitle

\section{Introduction}

Basic aim of the Workshop \cite{charm}
organised at Cern by U. Heinz
and C. Lourenco in December 1999 was the discussion of two questions
concerning open charm production in nucleus--nucleus (A+A) collisions
at Cern SPS energies:
\begin{itemize}
\item
do we want to measure it?
\item
can we measure it?
\end{itemize}
The workshop should be seen as an important element of an ongoing
debate on future of heavy ion physics at Cern SPS.
Two proposals were formulated for the experimental studies
of A+A collisions in year 2000 and beyond:
\begin{itemize}
\item
a measurement of open charm production at the top SPS energies
(158--200 A$\cdot$GeV) by NA60 \cite{na60} and NA49 \cite{add5} and
\item
a study of onset of deconfiment at low SPS energies 
(20, 30 and 80 A$\cdot$GeV) by NA49 \cite{add7}. 
\end{itemize}
The  Workshop allowed for a review and critical
discussion of theoretical and experimental
aspects of open charm proposals.

During the last 25 years a pQCD--based framework for a treatment 
of open 
(charm mesons and hyperons)
and hidden (charmonia)
charm production was formulated and it dominated the interpretation of
the experimental data.
It is based on the assumption that the perturbative QCD apparatus
is applicable for open charm production in 
hadronic and nuclear collisions and that charmonium creation
is due to binding interaction between 
perturbatively produced $c$--$\overline{c}$
quarks.
Since recent analysis of
$J/\psi$ and dimuon production in Pb+Pb
collisions 
put the standard framework to question, a new and vivid
discussion has started on the subject.
Several non--standard models for open and hidden charm
production were proposed,
the most controversial among them are statistical models.
During the Workshop  data and models were discussed
mainly
within the standard framework.
There are, however,
many excellent reviews 
\cite{std}
written by experts in this subject.
Thus in my brief report I will focus on the
new developments instead of repeating details of a well known
standard approach.
For completeness I will mention  recent results
which were not presented during the Workshop.

In section 2 of this report $J/\psi$ production and its
relation to open charm are discussed.
A summary of various approaches to open charm production
is given in section 3.
Finally a first attempts to measure an open charm yield in
A+A collisions at SPS and future plans for continuation of this 
effort are presented.

\section{$J/\psi$ Production}

The Workshop was opened by  Satz \cite{satz} who 
discussed   $J/\psi$ production and its relation to
open charm in nuclear collisions. 
The $J/\psi$ has been measured in A+A collisions over the last
15 years by NA38 and  NA50 Collaborations. 
This 
experimental program was mainly motivated by a hypothesis
of Matsui and Satz \cite{Ma:86} formulated within standard framework.
They suggested that $J/\psi$  may serve as a probe
of the state of matter created in the early stage of the collision.
Recent results and their new interpretations indicate that
Matsui--Satz approach may be incorrect.

Two years ago Gorenstein and myself \cite{Go:99,Ga:99} observed that 
the $J/\psi$ yield in nuclear collisions is proportional to 
the pion
multiplicity.
We further showed \cite{stat} that the data may be explained assuming
statistical production of $J/\psi$ at hadronization.
Recently several models in which $J/\psi$ creation
is due to coalescence of $c$--$\overline{c}$ quarks or 
$D$--$\overline{D}$ mesons at different stages of collision 
were formulated.
A brief summary of these approaches is given below.

\subsection{Standard Approach}

In the standard approach reviewd by Satz \cite{satz}
charmonium production is considered  as a three
stage process: 
the creation of a $c$--$\overline{c}$ pair,   formation of
a bound $c$--$\overline{c}$ state, and its subsequent interaction
with the surrounding matter.
The first process is calculated within pQCD, whereas 
non--perturbative QCD is needed to describe the last two stages.
The interaction of the $c$--$\overline{c}$ bound state with matter
results in suppression of the final $J/\psi$ yield
in comparison with its initial number.
In order to comapre the model to the data, the assumption
is made that the initial $J/\psi$ yield
is proportional to multiplicity
of Drell--Yan pairs.
Within this framework the significant suppression of $J/\psi$ 
production
relative to Drell--Yan which is observed
when going from peripheral to central Pb+Pb interactions
at 158 A$\cdot$GeV \cite{na50} 
is attributed to the formation of a QGP
in the latter collisions \cite{satz_rev}.  

The data on $J/\psi$ production in proton--nucleus (p+A) interactions
serve as a first test of the model. Aichelin \cite{aichelin} 
pointed out that
recent precise results on $x_F$ dependence of the charmonium
yield in p+A collisions \cite{pa}
are difficult to understand within
the standard framework. 

\subsection{Statistical Approach}

The statistical approach, presented
by Gorenstein and myself \cite{gazdzicki}, 
assumes
that $J/\psi$ mesons are created at hadronization
according to the available hadronic phase--space.
Thus, within this model, the $J/\psi$ yield is independent
of the open charm yield.
Moreover, the $J/\psi$ production is,   in good approximation,
insensitive to the state of matter created at the early stage. 
This is because charmonium is  created only at
the hadronization stage.
The model offers natural explanation of the proportionality
of the $J/\psi$ and pion yields and the magnitude of the
multiplicity of $J/\psi$ mesons in hadronic and nuclear collisions
\cite{stat}.
     
It should be noted that  a consistent picture of hadron production 
within this model is still
missing. 
As an example one can consider $\psi'$ production.
The statistical approach used for $J/\psi$ description works
for  $\psi'$ production in central A+A collisions \cite{Sh:97, Br:00}.
It does not give, however, a natural explanation of a substantial
increase of the $\psi'/J/\psi$ ratio measured in p+p and p+A interactions.

\subsection{Microscopical Coalescence Model}

Microscopical coalescence model \cite{micor}, 
presented by L\'evai \cite{levai},
assumes that $J/\psi$ mesons are formed at hadronization
as a result of coalescence of $c$--$\overline{c}$ quarks created
in the earlier stages of the collision.
By introduction of microscopic coalescence factors and accounting
for quark number conservation one can relate $J/\psi$ and open
charm production. 
Starting from the measured multiplicity of $J/\psi$ mesons in
central Pb+Pb collisions at SPS, one predicts mean
multiplicity of $c$--$\overline{c}$ pairs, $n_{c \overline{c}} \approx 3$.

\subsection{Statistical Coalescence Model}

Also in the 
statistical coalescence model,
recently introduced by Braun--Munzinger and
Stachel \cite{Br:00}, charmonium states are  produced 
at hadronization as a coalescence of earlier created
$c$--$\overline{c}$ quarks.
It is further assumed that the number of $c$--$\overline{c}$
quarks is given by pQCD and that the redistribution of these
quarks among hidden and open charm hadrons follows the 
maximum entropy
principle.
The model reproduces the measured $J/\psi$ yield in central
Pb+Pb collisions at 158 A$\cdot$GeV using parameters fitted
to the light hadron sector.

This model, however, does not give a natural explanation
for the observed proportionality of $J/\psi$ and pion yields
in nuclear collisions.

\subsection{Coalescence in QGP}

A possible contribution  from the coalescence of 
$c$--$\overline{c}$ quarks originating from different
elementary interactions was calculated by 
Thews, Schroedter and Rafelski \cite{Ra:00}.
The coalescence process is assumed to take place in  
the QGP phase in parallel with the 
expected disintegration of
charmonium states.
Starting from the pQCD charm yield
they calculated 
that the
coalescence contribution can lead to enhanced
production of $J/\psi$ mesons 
in A+A collisions at RHIC and LHC energies.
This prediction should be confronted with the
expectation  of
$J/\psi$ suppression derived within the standard
approach \cite{satz_rev}. 

\subsection{Secondary Hadronic Production}

Charmonium production resulting from the interaction
of $D$ and $\overline{D}$ mesons in hadronic matter
was considered by Braun--Munzinger and Redlich \cite{Re:00}.
The contribution from this process to the total 
$J/\psi$ yield was found to be small
even at LHC energies.  
This is not the case for $\psi'$ production, where
the secondary production may substantially exceed the
primary yield.

\vspace{0.2cm}
\noindent
Presented models predict very different relations
between hidden and open charm production.
In the statistical model the $J/\psi$ and open charm
yields are
independent.
Initial charmonium production is proportional to
$n_{c \overline{c}}$ in the standard approach. 
This relation is however modified by the suppression processes.
In the coalescence models  $J/\psi$ multiplicity
increases approximately as $n_{c \overline{c}}^2$.

It is therefore clear that data on open charm 
in A+A collisions are necessary in order to understand
mechanism of $J/\psi$ production.

\section{Open Charm Production}

Open charm production is not measured in A+A collisions.
The presentation of Capelli \cite{capelli} was, however, devoted
to the question whether open charm yield can be indirectly
estimated from the current data. 
NA50 observes enhancement of dimuon pairs in the mass
region between $\phi$ and $J/\psi$ peaks (intermediate mass
region, IMR) in central A+A collisions at SPS \cite{dimuon}.
A possible interpretation of this effect, favoured by NA50,
is that it is due to enhanced open charm production.
The enhancement factor is relative to the pQCD prediction.
Note that the possible enhamcement already questions the
validity of the standard approach \cite{std}.

The basic ideas concerning open charm production presented during 
the Workshop are summarised below.

\subsection{Standard Approach}

In standard model, revied by Vogt \cite{vogt},
one assumes validity of pQCD calculations
for open charm production in hadronic and nuclear interactions.
A 
comparison with the open charm
measurements in hadronic interactions serves as a first test
of the model.
In fact the data can be described assuming reasonable values
of input parameters (charm quark mass, $\Lambda_{QCD}$,
renormalization and factorisation scales).
One should note, however, that uncertainties in this parameters
lead to the variation of the calculated open charm
multiplicity by a  factor of about  100 (Frixione et al. \cite{std}). 

The model with the parameters fitted to the hadronic data
can be used to predict open charm multiplicity in p+A and A+A
collisions. 
Unfortunately present p+A data are too sparse for a precise
test of the model \cite{Ga:99}.

In good approximation the following $A$--dependence is expected 
for central A+A collisions:
\begin{equation}
\langle n_{c\overline{c}} \rangle_{AA} 
= \langle n_{c\overline{c}} \rangle_{NN} \cdot A^{4/3}.
\end{equation}
This results in an  estimate of the 
mean multiplicity of $D^0 + \overline{D}^0$ mesons
in central Pb+Pb collisions at 158 A$\cdot$GeV:
$\langle D^0 + \overline{D}^0 \rangle 
\approx 2 \cdot 10^{-1}$ \cite{Ma:99}. 

\subsection{Secondary Production}

The pQCD based calculations on
open charm production from secondary collisions occurring
in the expanding QGP were reported by Braun--Munzinger \cite{braun}.    
This additional contribution to the initial open charm
yield was found to be negligible at Cern SPS energies.

\subsection{Statistical Approach}

The statistical model, introduced by Gorenstein
and myself \cite{Go:99}, assumes that 
charm quarks and antiquarks are created according to
the early stage partonic phase space.
This non--standard assumption was motivated by a success
the statistical model of the early stage of
A+A collisions
in description of strangeness and entropy production.
Within the model  the expected
number of $c$--$\overline{c}$ pairs is about 8 in central
Pb+Pb collisions at 158 A$\cdot$GeV and the corresponding
number of $D^0 + \overline{D}^0$ mesons is about 6.

For central collisions of large enough nuclei ($A > 30$)
the grand canonical approximation can be used and consequently
one expects  that open charm multiplicity increases  as $A^1$.

\vspace{0.2cm}
\noindent
The pQCD based and statistical estimates of open charm
yield in central Pb+Pb collisions at 158 A$\cdot$GeV
differ by a factor of about 30.
The predicted $A$--dependence in these two approaches is also
 different.
Only direct measurements of open charm production in A+A
collisions can uniquely distinguish between these spectacularly 
different models.
In particular open charm measurements should give a unique
opportunity  to set limits to the applicability
of pQCD and statistical models of strong interactions.

\section{Open Charm Measurements}

\subsection{Present}  

\underline{NA50} (Capelli \cite{capelli}) 
reported increased production of dimuon pairs
in the intermediate mass region over the standard
sources in A+A collisions.
The enhancement factor is about 3 for central Pb+Pb
collisions at 158 A$\cdot$GeV.
The observed effect can be attributed to 
enhanced production of open
charm.

However other interpretations are also possible.
Wang argued \cite{wang} that
the data can be reproduced assuming that only the momentum
spectrum of charm  hadrons is modified in A+A collisions
with respect to p+p and p+A interactions.
Thermal radiation of dimuons was also shown to reproduce
the experimental results on dimuon enhancement \cite{dimuon}.

Finally one can argue \cite{bkg} that the background subtraction
procedure used by NA50 may lead to biased results 
in the case of central Pb+Pb collisions.

\vspace{0.2cm}
\noindent
\underline{NA49} \cite{gazdzicki} 
made first attempt to estimate an upper limit
of mean multiplicity of $D$ and $\overline{D}$ mesons
in central Pb+Pb collisions at 158 A$\cdot$GeV by a direct
measurement.
In this case invariant mass distribution of identified kaons
and pions is studied \cite{Ma:99}.
Using the currently analysed number of events ($4\cdot10^5$)
the estimated upper limit is on the level of the statistical
model prediction.

\subsection{Future}

\underline{NA49} \cite{add5}
is planning to significantly increase 
(up to $5\cdot10^6$ events) the
statistics of central Pb+Pb collisions at 158  A$\cdot$GeV
during this year's Pb--run. 
Statistical resolution of this new data should allow to
exclude models predicting 
$\langle D^0 + \overline{D}^0 \rangle > 1$.

\vspace{0.2cm}
\noindent
Shahoian \cite{shahoian} reported that
one of the main goals of the recently proposed and approved
experiment \underline{NA60} \cite{na60} is to study
the origin of the dimuon enhancement in the IMR observed by NA50.
Main components of NA60 experiment are:
\begin{itemize}
\item the present NA50 detector with the muon arm as a
basic instrument,
\item
a new pixel vertex spectrometer and
\item
a new beam scope.
\end{itemize}
This set--up will allow for a precise measurements
of the track impact parameter at the interaction point.
It will also result in significant improvement of the momentum
resolution in comparison to that achieved by NA50.
Consequently NA60 will be able to distinguish between
prompt dimuons (Drell--Yan and thermal contributions)
and dimuons from decays of charm hadrons.
This  should
lead to  clarification of the origin of dimuon enhancement.
NA60 was approved for p--beam in 2001.
The decision concerning requested ion beams in 2002 and 2003
will be made in September this year.

\vspace{0.2cm}

Systematic charm measurements in nuclear collisions 
are key elements in the upcoming RHIC and LHC
programs.
Indirect and direct measurements of open charm should be possible
in \underline{PHENIX} at RHIC (Averbeck \cite{averbeck}) and in 
\underline{ALICE} at LHC (Safarik).

\ack

I would like  thank Marco van Leeuwen for comments
to the manuscript.

\end{document}